# IRRADIATION OF MATERIALS

# WITH SHORT, INTENSE ION PULSES AT NDCX-II*

P.A. Seidl[†], Q. Ji, A. Persaud, E. Feinberg[*], B. Ludewigt, M. Silverman, A. Sulyman,

W.L. Waldron, T. Schenkel,

Lawrence Berkeley National Laboratory, Berkeley, USA

J.J. Barnard, A. Friedman, D.P. Grote,

Lawrence Livermore National Laboratory, Livermore, USA

E.P. Gilson, I.D. Kaganovich, A.D. Stepanov,

Princeton Plasma Physics Laboratory, Princeton, USA

F. Treffert, M. Zimmer, TU Darmstadt, Darmstadt, Germany

*Corresponding author :*
  Peter Seidl
  Lawrence Berkeley National Laboratory
  1 Cyclotron Road
  Mailstop 58-0111
  Berkeley, CA 94720
  Telephone : +1(510)486-7653
  PASeidl@lbl.gov

  # of manuscript pages: 15
  # of tables: 0
  # of figures: 7

---

[*] Now XTD-IDA, Los Alamos National Laboratory, Los Alamos, NM, USA

# Abstract

We present an overview of the performance of the Neutralized Drift Compression Experiment-II (NDCX-II) accelerator at Berkeley Lab, and report on recent target experiments on beam driven melting and transmission ion energy loss measurements with nanosecond and millimeter-scale ion beam pulses and thin tin foils. Bunches with around $10^{11}$ ions, 1-mm radius, and 2-30 ns FWHM duration have been created with corresponding fluences in the range of 0.1 to 0.7 J/cm$^2$. To achieve these short pulse durations and mm-scale focal spot radii, the 1.1 MeV He$^+$ ion beam is neutralized in a drift compression section, which removes the space charge defocusing effect during final compression and focusing. The beam space charge and drift compression techniques resemble necessary beam conditions and manipulations in heavy ion inertial fusion accelerators. Quantitative comparison of detailed particle-in-cell simulations with the experiment play an important role in optimizing accelerator performance.

# INTRODUCTION

Intense pulses of ions in the MeV range enable new studies of the properties of matter ranging from low intensity (negligible heating, but active collective effects due to proximate ion trajectories in time and space), to high intensity where the target may be heated to the few-eV range and beyond. By choosing the ion mass and kinetic energy to be near the Bragg peak, dE/dx is maximized and a thin target may be heated with high uniformity (Grisham, 2004). The Neutralized Drift Compression Experiment (NDCX-II) was designed with this motivation (Friedman *et al.*, 2010; Barnard *et al.*, 2014; Waldron *et al.*, 2014).

Reproducible ion pulses ($N>10^{11}$ /bunch), with bunch duration and spot size in the nanosecond and millimeter range, meet the requirements to explore the physics topics identified above. The formation of the bunches generally involves an accelerator beam with high perveance and low emittance, attractive for exploring basic beam physics of general interest, and relevant to the high-current, high-intensity ion beams needed for heavy-ion-driven inertial fusion energy (Bangerter *et al.*, 2013; Davidson *et al.*, 2002).

Furthermore, short ion pulses at high intensity (but below melting) enable pump-probe experiments that explore the dynamics of radiation-induced defects in materials. While progress on accessing multi-scale dynamics is being made (Wallace *et al.*, 2016), reliable experimental benchmarking of defect dynamics on short time scales is currently lacking. For high peak currents and short ion pulses, the response of the material to radiation may enter a regime of overlapping collision cascades initiated by the incident ions. Effects of varying ion dose rates may be transient (no memory effect at a subsequent pulse) or lead to enhanced defect accumulations and the short, intense pulses of ions provide an opportunity to observe the time-resolved multi-scale dynamics of radiation-induced defects (Schenkel *et al.*, 2013; Persaud *et al.*, 2015; Bai *et al.*, 2010). For example, one diagnostic is a transmission energy loss measurement with thin (~1 micron) for incident 1-MeV $He^+$ ions) on single crystal targets. The time-resolved measurements can provide insights and constraints on models of defect formation, recombination and diffusion where high impact application areas include radiation effects in structural materials for fission and future fusion reactors.

Topics pertinent to materials in fusion reactors include radiation damage and void swelling of structural components in magnetic confinement systems. Intermittent bursts of intense fluxes on plasma-facing components can arise with the emergence of edge-localized modes (ELM's) and possible disruptions. Inertial confinement fusion reactors are subject to pulsed radiation fields, with relevant timescales ranging from nanoseconds (x-rays from the target plasma heated by the charged fusion reaction products) to microseconds (charged particles from the primary fusion reactions and other ions from the destruction of the target). The variety of designs of the chamber wall range from dry walls (the structural components must absorb the full flux) to wetted walls (thin liquids on the surface) and thick liquid (flowing jets that absorb the neutrons and other radiation). Repetitive thermal cycling (at ~10 Hz) of structural materials below and above the recrystallization temperature are characteristic of IFE reactor chambers and the pulsed intense beams described below offer opportunities to probe the relevant scientific questions.

* PASeidl@lbl.gov Work supported by the US DOE under contracts DE-AC0205CH11231 (LBNL), DE-AC52- 07NA27344 (LLNL) and DE-AC02-09CH11466 (PPPL).

For the accelerator performance already achieved, these are promising research topics for NDCX-II, in contrast to driving of materials into warm dense matter states with temperatures >1 eV, which require a beam fluence greater than 5 J/cm$^2$ with compression of about $3 \times 10^{11}$ ions to within 1 ns pulses and 1-mm$^2$ beam spots (Barnard *et al*, 2014). The present performance of NDCX-II falls short of requirements to reach warm dense matter states by a factor of four in ions/pulse, and factors of two in pulse length and beam spot size, respectively. Identifying and correcting performance limiting factors (including ion beam emittance, chromatic and geometric aberrations in the focusing magnets, and compression and acceleration waveform optimization) can lead to further increases of beam performance from the current 1 A/mm$^2$ scale limit towards tens of amperes of peak current.

## ACCELERATOR PERFORMANCE AND BEAM MODELING

Having switched from our earlier installed surface emission type ion source for lithium, our multi-cusp, multiple-aperture plasma ion source is capable of generating high purity ion beams of, for example, protons, helium, neon and argon (Ji *et al.*, 2016). To date, we have used the source solely for the generation of He$^+$ ions. Furthermore, helium at about 1 MeV is nearly ideal for highly uniform volumetric energy deposition, because particles enter thin targets slightly above the Bragg peak energy and exit below it, leading to energy loss in the target, uniform within several percent.

Ion induction accelerators can simultaneously accelerate and rapidly compress beam pulses by adjusting the slope and amplitude of the voltage waveforms in each acceleration gap. In NDCX-II, this is accomplished with 12 compression and acceleration waveforms driven with peak voltages ranging from 15 kV to 200 kV and durations of 0.07-1 µs (Seidl *et al*, 2015). The first seven acceleration cells are driven by spark-gap switched, lumped element circuits tuned to produce the required cell voltage waveforms. These "compression" waveforms induce a head-to-tail velocity ramp because of their characteristic triangular shape, and have peak voltages ranging from 20 kV to 50 kV. An essential design objective of the compression pulsers is to compress the bunch from about 0.6 µs to <70 ns so that it can be further accelerated and bunched by the 200-kV Blumlein pulsers which drive the last five acceleration cells.

In the final drift section, the bunch has a head-to-tail velocity ramp that further compresses the beam by an order-of-magnitude. The space-charge forces are sufficiently high at this stage to require that dense plasma ($n_{plasma} > n_{beam}$), generated prior to the passage of the beam pulse, neutralizes the beam self-field and enables focusing and bunching of the beam to the millimeter and nanosecond range (Gilson *et al*., 2013; Gilson *et al*., 2012). The beam perveance:

$$Q = \frac{2qI}{4\pi\varepsilon_o M v^3}$$

is high throughout the accelerator (~2x10$^{-3}$) at the first acceleration gap and higher as the beam compresses ($q$ is the ion charge, $I$ is the beam current, $M$ is the ion mass and $v$ is the ion velocity). When passing through the neutralizing plasma the coasting beam compresses with negligible space-charge repulsion, thus the spot size and bunch duration are limited mainly by voltage waveform fidelity, chromatic aberrations and emittance of the beam.

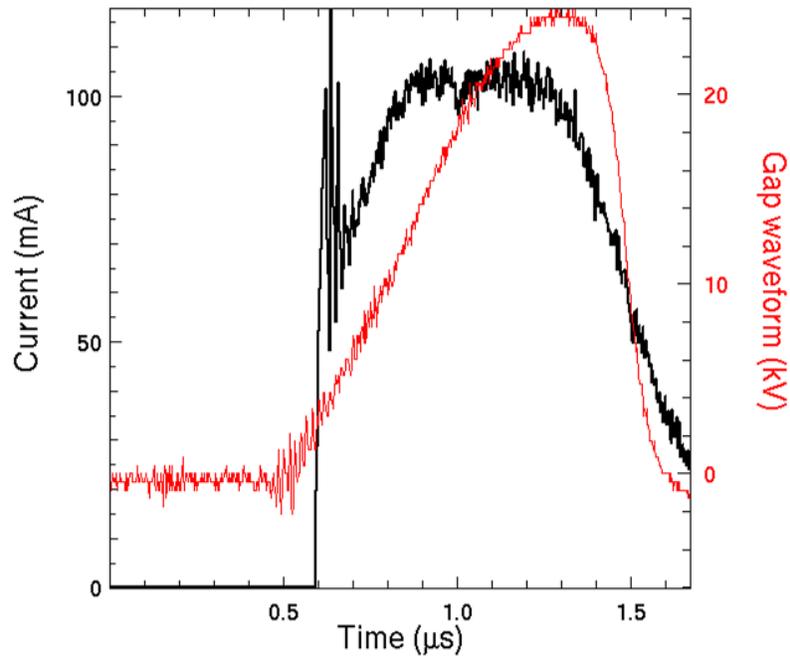

Figure 1: Warp PIC simulations are initialized with measured acceleration gap waveforms as shown in the example above, for z=1003 mm downstream of the ion source, at the middle of the second compression cell acceleration gap.

The NDCX-II accelerator provides a platform to extend the limits of intense beam and beam-plasma physics. For example, an ion-electron two-stream instability has been predicted (Tokluoglu *et al.*, 2015), and it may be observed in NDCX-II by passing a nearly constant energy beam through a plasma. The manifestation would be a significant transverse defocusing and longitudinal bunching of the specially prepared beam. In NDCX-II, the effect is normally absent because of the imposed velocity ramp on the beam distribution.

Another interesting beam physics opportunity is to demonstrate the collective focusing of an ion beam in a weak magnetic field (Dorf *et al.*, 2012). The focusing occurs due to the formation of a strong radial self-electric field due to the rearrangement of the plasma electrons moving with the beam, and in response to a weak magnetic field. The magnetic field is established by the final solenoid near the end of the neutralized drift section. But, instead of using the full strength of the final solenoid (5-8 Tesla) to focus the beam, equivalent focusing may be achieved with only ~0.02 Tesla. If demonstrated to be practical, the focusing magnet strength would be greatly reduced with a corresponding reduction of the stray magnetic field at the nearby target plane – as preferred for some beam-target interaction experiments.

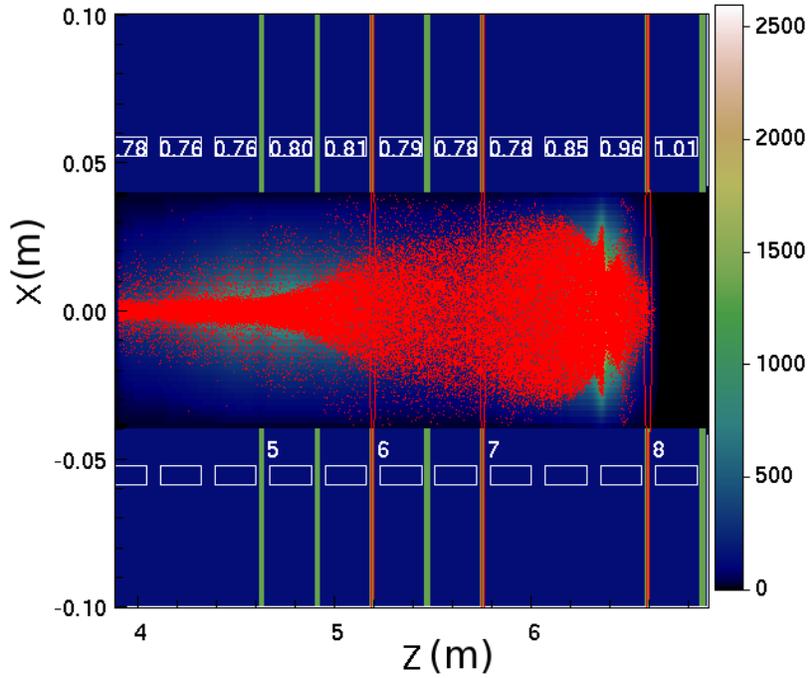

Figure 2: The simulations are initialized with measured magnetic field strengths, as shown in the particle distribution at t = 2660 ns. The x-z projections of the beam distribution show the main features of the particle transverse distribution as well as halo particle loss. The red contour lines show the equipotential from the applied acceleration fields. The units of x and z are in meters. The white labels within rectangles indicate the solenoid field strength in Tesla. The intensity bar (right) indicates the electric potential (Volts) due to the beam space charge.

We have enhanced the integration of simulations with the experimental data acquisition via automatic import from the data acquisition database of injector and acceleration waveforms from the experiment, along with all of the particular focusing magnet settings. Thus, the influence of small imperfections (timing, waveform shape) in the acceleration waveforms is included in the simulations to provide more accurate predictions of the accelerator performance. Figure 1 shows the measured acceleration voltage for a particular shot applied to the acceleration gap. The Warp (Friedman *et al.*, 2010) simulation also shows the arrival time of the beam pulse at the same gap – valuable for evaluating the effects of pulser timing fine adjustments. In Fig. 2 the particle distribution is shown for the same pulse after passage through several solenoids (downstream of Fig. 1). We have demonstrated that the simulations can keep up with the accelerator repetition rate of ~1/min with parallel runs running on a handful of processors.

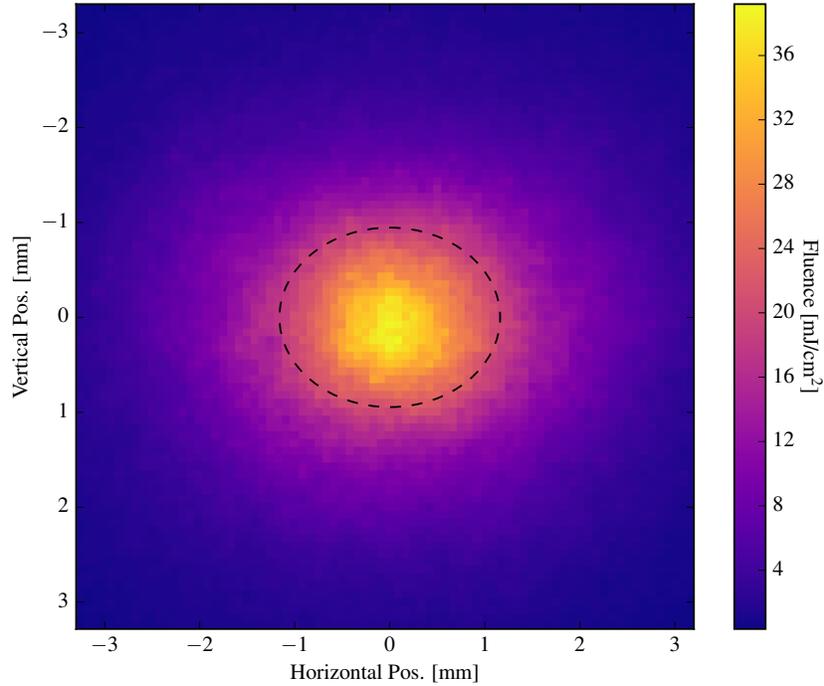

Figure 3: Beam intensity imaged by an intensified CCD camera. An $Al_2O_3$ scintillator intercepts the beam at the target plane. The dotted line shows the beam profile with 2 mm FWHM.

Figure 3 shows the beam current density profile measured with a scintillator and gated CCD camera for a pulse with $2.5 \times 10^{10}$ $He^+$ ions (1 MeV) and 6 ns FWHM. The FWHM beam radius is ~1 mm. The envelope and the beam current vary significantly through the pulse and solenoid focusing is mostly balancing the transverse space charge. The magnetic fields are chosen to minimize envelope mismatch and transmit the highest charge to the target plane with the best focal properties. Simulations suggest that the total bunch charge can be increased several fold, via adjustments to the solenoid and compression voltage waveforms. Using scintillator beam imaging at various locations in the accelerator and the advances in modelling described above to infer the beam emittance, we are currently exploring the effects on the focused intensity at the target due to the initial beam emittance, emittance growth, chromatic effects and waveform fidelity.

We have installed scintillator beam imaging diagnostics at three locations in the accelerator and measured the beam transverse density distribution while parametrically varying the solenoid field strengths. The results were used as input to models of the beam and iterated to improve the beam focusing properties at the entrance to the drift compression section. As a result, we have recently doubled the peak ion current delivered to the target to >2 Amperes of He$^+$, as shown in Fig. 4. The current profile shows a sharp peak followed by a longer tail. After the first 5 ns, when the narrow current spike is at ~12% of its peak value (2.1 Amperes in this case), the integrated dose is 37% of total. After 20 ns, about 87% of the charge is accumulated. The later arriving particles are due to compression cell waveform shape and timing imperfections. Here, we had injected a 0.8-μs pulse with a total charge of 65 nC. The transmitted charge to the target plane to date is thus approximately 20% and we are investigating limiting factors. Beam diagnostics suggest significant particle loss at injection and also after the last acceleration cell. At the current performance level the total ion pulse energy is 12 mJ or 0.15 J/cm$^2$ based on the Faraday cup and scintillator beam images. The highest intensity we have achieved to date is 0.7 J/cm$^2$ for slightly longer pulses with more charge and tighter focus (Seidl *et al.*, 2016).

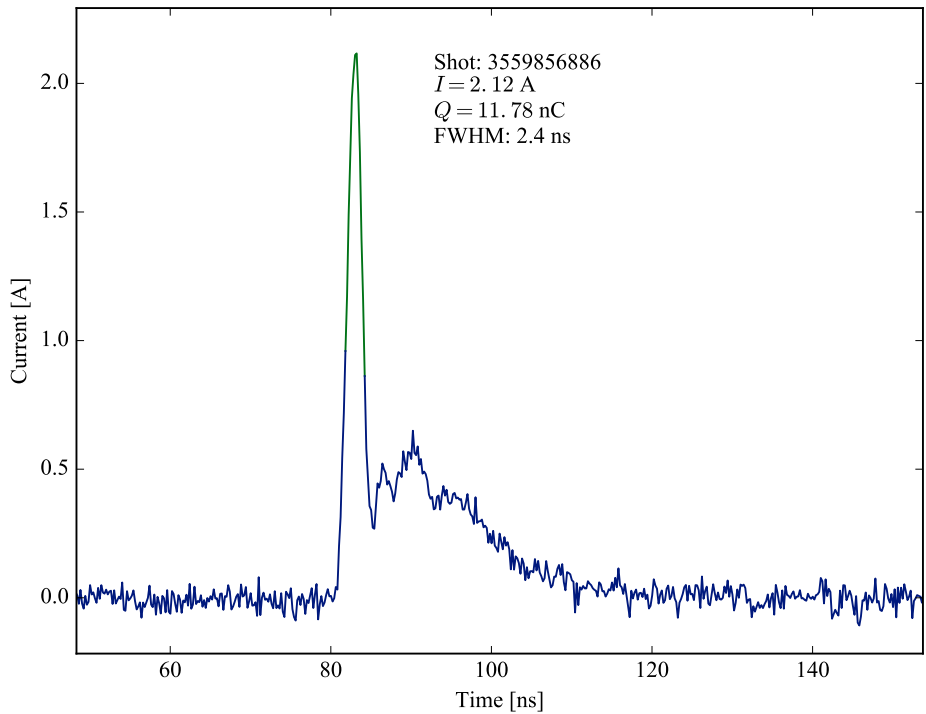

Figure 4: The Faraday cup response of the 1.1–MeV He$^+$ beam current shows a peak of 2.1 Amperes and an integrated charge of 12 nC.

# TARGET EXPERIMENTS

Figure 5 shows an example of target heating with intense ion pulses of a few nanosecond duration, inspected ex situ under an optical microscope. The 15 mm-diameter target is remotely moveable, thus allowing several shots per target on non-irradiated regions. The target thickness was 0.3 μm. The ion pulse was similar to the data shown in Figure 4, 1 MeV He$^+$, 2 A peak current, 2.4 ns FWHM in a spot with radius of 1 mm. We observe surface morphology changes on the tin foil that are indicative of melting and also cracks from rupturing (likely due to thermal stress). This degree of damage is consistent with the (relatively low) melting point of tin (505 K), the latent heat of fusion (melting) of 58.5 J/g and the energy deposited by the ion pulse of about 0.4 mJ/cm$^2$. We are presently exploring the dynamics of the disruptions due to rapid heating using a 3D multi-physics multi-material code including the effect of surface tension (Liu *et al.*, 2017).

In order to measure ion energy loss in materials, including while driving phase-transitions during the ion pulse, we have implemented a time-of-flight (TOF) ion transmission energy loss capability. Prior to, and following target shots, we measure the temporal profile of the beam with a fast Faraday cup (0.4 nanosecond time resolution, Figure 4) (Sefkow *et al.*, 2006). There is some jitter in the arrival time of the beam due to the timing accuracy of the twelve high-voltage pulsers that are part of the ion acceleration sequence, corresponding to an 18 keV variation in the beam energy. Thus, to monitor and correct for this arrival time variation, the thin foil targets and holder are electrically isolated and the electrical signal generated by the ions striking the target is recorded. We then use this pulse to start the TOF measurement. The transmitted ions are then recorded in a second fast Faraday cup 46 cm downstream of the target. This second cup is a variation of a design used at the target plane: It has a single (vs. two) hole plate 1 mm from the collector which mainly serves to establish a short electrostatic boundary and in turn an ion transit time and time resolution <1 ns. Results of the TOF transmission ion energy loss measurements are shown in Fig. 6, for 0.944 MeV He$^+$ incident on a 0.54-μm thick Sn foil (thickness measured with Rutherford backscattering). The measured flight time of about 80 ns translates into a peak energy loss of 230 ± 14 keV, where the uncertainty is the rms shot-to-shot variation in the arrival time at the peak of the ion current pulse. This result agrees with SRIM (Ziegler) simulations result of 234 keV, with an associated energy spread due to straggling of ±13 keV. The uncertainty expected from the foil thickness is ±1% (including small surface contamination layers). We note that our 1 MeV He$^+$ pulses implement Bragg-peak heating (Grisham, 2004) with uniform energy deposition across the foil within 2% (SRIM). In our TOF experiments the straggling energy spread is convolved with ±25 keV energy variation due to neutralized drift compression chromatic effect.

Flight time increases for a series of shots with varying fluences (15±1, 40±2, and 153±9 mJ/cm$^2$) show a slight variation with fluence (12-keV higher at low fluence), but close to the rms variation in ion kinetic energy in repeated shots. We will study this further in future experiments. The fluences were varied by defocusing the beam using the final focus magnet while the ion pulse length remained constant. The pulses in Figure 7 show beam profiles after transmission of a foil target. Though we did not observe intensity dependent effects in these measurements and the results demonstrate a transmission TOF ion energy loss measurement capability. Electronic energy loss and electronic excitations form short, intense ion pulses can lead to transient populations conduction electrons in semiconductors and insulators, which can increase energy loss rates. For pulses as shown in Figure 4, we estimate a conduction band population at up to 10% of atomic density and we are now exploring this hypothesis.

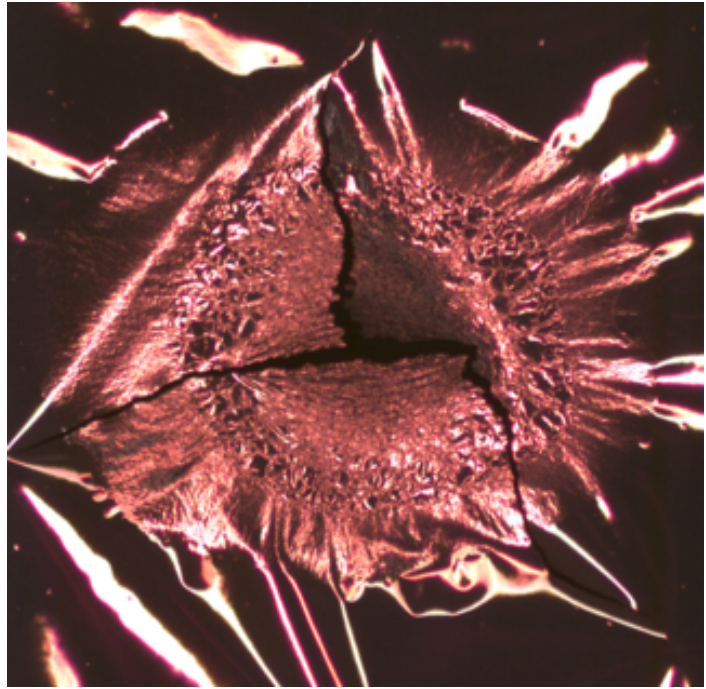

Figure 5: A microscope image of a 0.3-μm Sn foil after being struck by one helium beam pulse (40 mJ/cm$^2$, 1 MeV) shows surface structures from melting and re-solidification and cracks from rupturing. The field of view is ≈ 6mm.

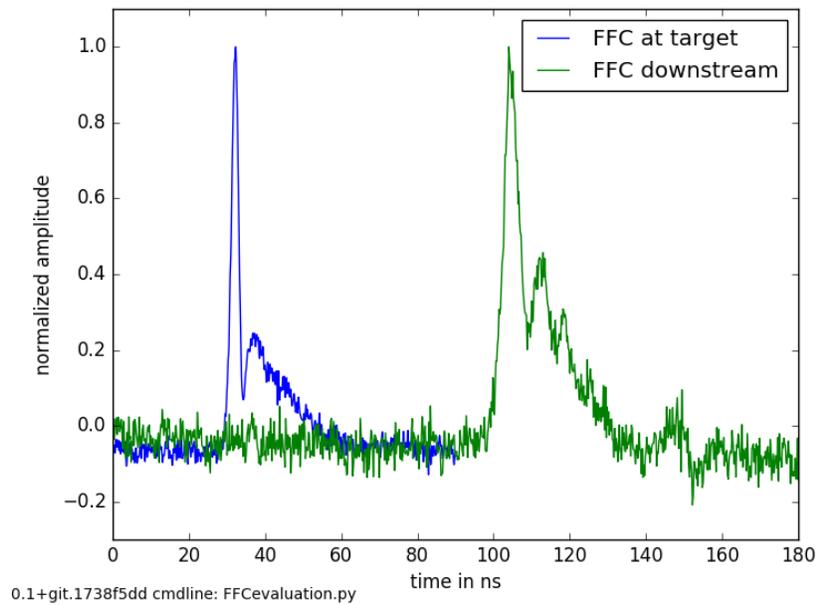

Figure 6: The Faraday cup at the target plane and at the downstream target plane show the beam current waveform for the incident and transmitted beam.

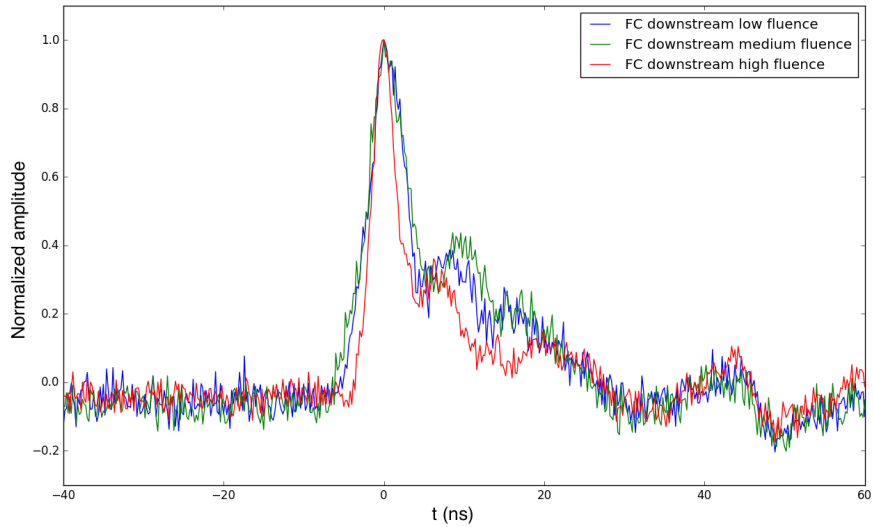

Figure 7: The fluence of the transmitted ions as measured with the fast Faraday cup 47 cm downstream of the target.

Thermal annealing of ion implanted materials often takes place on a timescale of milliseconds to seconds. Light ion beam driven annealing was explored in the 1980s with intense light ion beams and 100 ns long pulses (Baglin *et al*, 1981). With NDCX-II, we now have an opportunity to locally excite materials with intense ion pulses on a timescale of 1 to 5 ns. This enables driving of materials far from equilibrium to form desired materials phases that can then be stabilized by rapid quenching. This promises to allow materials development and optimization beyond limits of conventional thermal processing, for example for niche applications in sensing and quantum information processing (Schwartz, et al, 2014; Bienfait *et al.*, 2016).

## ACKNOWLEDGEMENTS

Work supported by the US DOE under contracts DE-AC0205CH11231 (LBNL), DE-AC52- 07NA27344 (LLNL) and DE-AC02-09CH11466 (PPPL). F. Treffert and M. Zimmer thank Prof. Markus Roth for stimulating discussions.